\documentstyle[prl,aps,multicol]{revtex}

\begin{document}
\draft


\title{Waves in Open Systems via Bi-orthogonal Basis}

\author{P. T. Leung${}^{(1)}$,  W. M. Suen${}^{(1,2)}$,
C. P. Sun${}^{(1,3)}$ and K. Young${}^{(1)}$}

\address{${}^{(1)}$Department of Physics, 
The Chinese University of Hong Kong, 
Hong Kong, China}

\address{${}^{(2)}$McDonnel Center for the Space Sciences, Department of Physics,\\
Washington University, St Louis, MO 63130, USA}

\address{${}^{(3)}$Institute of Theoretical Physics,
Academia Sinica, Beijing 100080, China}

\date{\today}
\maketitle

\begin{abstract}

Dissipative quantum systems are sometimes phenomenologically
described in terms
of a non-hermitian hamiltonian $H$, with
different left and right eigenvectors forming a bi-orthogonal basis.
It is shown that
the dynamics of waves in open systems can be cast exactly into this form,
thus providing a well-founded realization of the phenomenological
description and at the same time placing these open systems
into a well-known framework.  The formalism
leads to a generalization of norms and inner products
for open systems, which in contrast to earlier works
is finite without the need for regularization.
The inner product allows transcription of much of
the formalism for conservative systems, including
perturbation theory and second-quantization.

\end{abstract}

 
\begin{multicols}{2}

\noindent {\bf Introduction}

Dissipative systems can be discussed in many ways.  The fundamental
approach recognizes that energy flows from the system $S$ to a bath
$B$, whose degrees of freedom are then eliminated from the path
integral or equations of motion \cite{diss}.  While rigorous, this
approach is inevitably complicated, and often leads to
integro-differential equations for time evolution.  An alternate
phenomenological approach {\em postulates}\/ a non-hermitian
hamiltonian (NHH) $H$, whose left and right eigenvectors form a
bi-orthogonal basis (BB) \cite{wong,lath,rott,faisal,dattoli,sun93}.
These NHHs with discrete BBs can sometimes be obtained from a full
quantum theory, but usually under some approximations
\cite{faisal,others}.

This Letter discusses a class of models of waves in open systems.
These are scalar fields $\phi(x,t)$ in 1 d, described by the wave equation.
Outgoing wave boundary conditions cause the system to be dissipative.
We show that these open systems
are {\em exactly}\/ described by an NHH with a BB formed by the 
resonances or quasinormal modes (QNMs).
This connection on the one hand provides the phenomenological approach
with a realization which has an impeccable pedigree
rigorously traceable to the fundamental approach, and on the other
hand places earlier work on such open systems into a familiar framework.
A generalized inner product emerges;
in contrast to previous works, it is finite and
requires no regularization.  Under the generalized inner product,
the hamiltonian $H$ is symmetric, which opens the way to a clean
formulation of perturbation theory and second-quantization
in terms of the QNMs of the system.

\noindent {\bf Waves in Open Systems}

We consider waves in 1 d described by
$\left[ \rho(x) \partial_t^2-
\partial_x^2 \right] \phi (x,t) = 0$
on the half line $[0, \infty)$,
with $\phi(x=0,t)=0$ and $\phi(x,t)$ approaching zero
rapidly as $x \rightarrow \infty$ \cite{fn2}.
Let the system $S$ be the ``cavity" 
$I= [0,a]$, and the bath $B$ be
$(a, \infty )$, where $\rho(x) = 1$.  
Energy is exchanged between $S$ and $B$ only through the boundary $x=a$.
We impose the outgoing wave condition 
$\partial _t{\phi }(x,t)=-\partial _x \phi (x,t)$ for $x>a$.

This mathematical model is relevant for many physical systems:
the vibrations of a string with mass density $\rho$ \cite{string};
the scalar model of EM in an optical cavity
(the node at $x=0$ is a totally reflecting mirror,
and a partially transmitting mirror at $x=a$ can be modeled
by $\rho(x) = M\delta(x-a)$) \cite{lang}; or
gravitational radiation from a star with radius $a$
\cite{price}.
The wave equation can be mapped to the Klein-Gordon
equation with a potential $V(x)$ \cite{kg}, which is relevant for gravitational waves \cite{chand};
here $\phi$ is the perturbation about the spherical background
metric of a star,
$x$ is a radial coordinate related to the circumferential radius $r$, 
and $V$ describes the wave scattering by the 
background metric.
Gravitational waves carrying the signature of the QNMs
of black holes may soon
be observed by new detectors such as LIGO and VIRGO \cite{abram}.

For the ``cavity" $I =[0,a]$,
the outgoing condition is imposed at $x=a^+$
only.  The QNMs are factorized solutions on $I$:
$\phi(x,t) = f_n(x) e^{-i\omega_n t}$, with 
$ [\partial _x^2 + \rho (x) \omega _n^2] \, f_n(x) = 0$.
These are observed in the frequency domain
as resonances of finite width (e.g., the EM spectrum
seen outside an optical cavity) or in the time domain as
damped oscillations (e.g., the numerically
simulated gravitational wave signal from the vicinity of
a black hole).  It would obviously be interesting to be able
to describe these QNMs in a manner parallel to the normal
modes (NM) of a conservative system.

These QNMs form a complete set on $I$ if
(a) $\rho(x)$ has a discontinuity at $x=a$ to provide a natural
demarcation of the ``cavity", and (b) $\rho(x) = 1$ for $x>a$,
so that outgoing waves are not scattered back into
the system \cite{lly1}.  Under these conditions,
one can expand
$\phi(x,t)=\sum_n a_n f_n(x) e^{-i\omega _nt}$
for $x \in I$ and $t \ge 0$, thus allowing
an {\em exact} description of the
system in terms of {\em discrete}\/ variables
(modes spaced by $\Delta \omega \sim \pi /a$)
rather than a continuum.
Nevertheless, the analogy with conservative systems is
still not apparent: Is there a natural inner product (with which
to do projections and thus to prove the uniqueness of expansions)?
Is there a norm to scale wavefunctions (noting that
$f_n$ diverges at spatial infinity)?  Can perturbation theory be
formulated (noting that the usual proofs require an inner
product to define orthogonality)?
Can the theory be second-quantized?
This Letter shows that all these questions have
natural answers in the language of
a BB.

\noindent {\bf Phenomenological non-Hermitian Hamiltonians and Bi-orthogonal Bases}

Though not rigorously
founded upon a genuine quantum theory,
NHHs with BBs are nevertheless well developed as a {\em postulatory}\/
system \cite{wong,lath}.
Consider a space $W$ on which is defined
a non-hermitian operator $H$ and 
a conjugate linear duality transformation $D$:
$D \, \left( \alpha  |\Phi \rangle + \beta |\Psi \rangle \right)
=  \alpha^* D \, |\Phi \rangle + \beta^* D\, |\Psi \rangle$,
such that  
$DH = H^{\dagger}D$ \cite{fn1}.
The BB consists of the two set of
eigenvectors $|F_n \rangle \in W$ and
$|G_n\rangle = D\, |F_n \rangle \in {\tilde W} = D(W)$
satisfying
$H \, |F_n\rangle =\omega _n \, |F_n\rangle$,
$H^{\dagger}\, |G_n\rangle = \omega_n^* \, |G_n\rangle$,
where the two eigenvalues are related by duality.
By projecting the eigenvalue equations
on $\langle G_n |$ and $| F_n \rangle$, it follows easily that
$\langle G_n|F_m\rangle =0$, for $m \neq n$.

It is usually assumed that these eigenstates are complete, so that
any vector can be expanded as
$|\Phi \rangle = \sum_n a_n \, |F_n\rangle$,
with
$a_n = \langle G_n|\Phi \rangle / \langle G_n|F_n\rangle$,
leading immediately to the resolution of the identity and of the time-evolution
operator

\begin{eqnarray}
1 &=& \sum_n \frac{|F_n\rangle \langle G_n|}{\langle G_n|F_n\rangle}
\label{eq:comp1} \\
e^{-iHt} &=& \sum_n \frac{|F_n\rangle e^{-i\omega_nt}  \langle G_n|}
{\langle G_n|F_n\rangle} 
\label{eq:hrep}
\end{eqnarray}

\noindent
which in principle solves all the dynamics \cite{fn5}.

\noindent {\bf Bi-orthogonal Basis for the Wave Equation}

BBs are widely used in many disciplines, for example
in the theory of wavelets \cite{chu} and to describe
excited molecular systems \cite{rott,stanton}.
The left and right eigenvectors
of the Maxwell operator are typically used to represent
the Green's function for EM fields in open cavities
\cite{morse,cheng,mis}, or to evaluate
Fox-Li states \cite{fox}.  Here we seek a parallel with quantum mechanics,
similar to earlier works for generalized oscillators
\cite{sun} and 
the classical wave equation (without dissipation due to leakage) \cite{wu}.
The problem at hand, where there is dissipation
due to outgoing waves, was formulated in this manner recently \cite{twocomp},
and is briefly sketched below, especially as it relates to the BB.

It is natural to introduce the conjugate momentum
${\hat \phi} = \rho(x) \partial_t \phi$,
and the two-component vector
$|\Phi \rangle = (\phi, {\hat \phi } )^{\rm T}$.
In terms of this, the dynamics can be cast into the
Schr\"odinger equation with the NHH

\begin{equation}
H = i \pmatrix{0 & \rho (x)^{-1} \cr  \partial _x^2 & 0}
\label{eq:h}
\end{equation}

\noindent
The identification ${\hat \phi}=\rho \partial_t \phi$ follows
from the evolution equation \cite{bj}.

The natural definition of an inner product between
$|\Psi\rangle = (\psi, {\hat \psi})^{\rm T}$ and
$|\Phi\rangle = (\phi, {\hat \phi})^{\rm T}$
on $[0,\infty)$ is 

\begin{equation}
\langle \Psi | \Phi \rangle
= \int_0^{\infty} \left( \psi^* \phi + {\hat \psi}^*{\hat \phi} \right ) \, dx
\label{eq:inner1}
\end{equation}

\noindent
However, on account of the assumed asymptotic behavior,
the integral is convergent.

For outgoing waves, 
we consider only the space $U$
of such vectors $|\Phi\rangle$ defined on $[0, \infty)$
which satisfy the outgoing condition
${\hat \phi} = -\phi'$ for $x>a$.  The 
bath variables are eliminated 
simply but exactly by projecting 
to the space $W$ of vectors  $|\Phi\rangle$ defined
on $I$ and which satisfy ${\hat \phi} = -\phi'$ at $x=a^+$.
The QNMs are right-eigenvectors of $H$:
$|F_n\rangle \equiv 
(f_n , {\hat f}_n)^{\rm T} =
(f_n , -i\omega_n \rho f_n )^{\rm T}$.
The duality transformation is
$D \, ( \phi_1, \phi_2)^{\rm T} = -i ( \phi_2^* , \phi_1^* )^{\rm T}$.

For open systems, a crucial concept is the inner product between
one vector and the dual of another, to which we give a compact notation:

\begin{equation}
(\Psi, \Phi) \equiv \langle D \Psi | \Phi \rangle
= i\int_0^{\infty}  \left( {\hat \psi} \phi + \psi {\hat \phi} \right ) \, dx
\label{eq:inner2}
\end{equation}

\noindent
which is linear in both vectors, and cross-multiplies the two
components, properties to be emphasized below.
This bilinear map plays the role of the inner
product for conservative systems.

Our notation does not distinguish
between functions (say $|\Phi\rangle$)
defined on $[0, \infty)$ and their restrictions
to $I$; the former are in $U$ and the latter are in $W$,
with the association between them being many-to-one.
As written in (\ref{eq:inner2}), the inner product involves the wavefunctions
outside $I$, i.e., it appears to be defined on $U$ rather than $W$.
However, one can
completely eliminate the bath degrees of freedom:
because of the outgoing conditions, the
integrand on $(a, \infty)$ reduces to a total derivative, and (\ref{eq:inner2})
can be written purely in terms of the inside variables \cite{twocomp}:

\begin{equation}
(\Psi , \Phi) 
= i\left\{\int_0^{a^+} \left( {\hat \psi} \phi + \psi {\hat \phi} \right) \, dx
\,+\, \psi(a^+)\phi(a^+) \right\}
\label{eq:inner3}
\end{equation}

\noindent
The surface term is the only remnant of the outside.
Thus, (\ref{eq:inner3}) can be regarded as a bilinear map 
(or loosely an inner product) defined on $W$ \cite{twocomp}.  
The somewhat peculiar
structure (e.g., the cross-multiplication between the
two components and the appearance of the surface term)
is now seen to arise naturally from (\ref{eq:inner1})
upon the introduction of the duality transformation.
In the limit where the escape of the waves is small,
the generalized norm of an eigenvector $ (F_n , F_n) $
reduces to $2\omega_n$ times the conventional norm; this is the
reason for choosing the phase convention for $D$.
The ability to normalize QNM wavefunctions is nontrivial, since
$f_n$ diverges at spatial infinity, and a naive expression such
as $\int_0^{\infty} |f_n|^2 dx$ would not be
appropriate.

The diagonal version  $(\Phi, \Phi)$ for the special case of
QNMs was first introduced by Zeldovich \cite{zel}
in a form that involved (a) $\phi$ outside $I$ (so that
it is defined on $U$ rather than $W$) and
(b) regularization of the divergent integral rather than a surface term; it
was later re-cast into the form (\ref{eq:inner3}) and generalized
to 3 d and EM fields \cite{per1}.  The off-diagonal
form $(\Psi,\Phi)$ was later introduced \cite{twocomp}.
Here, by relating the discussion to bi-orthogonal states and the duality
transformation, it is seen that these concepts emerge naturally, including the specific
form of (\ref{eq:inner3}).

An inner product equivalent to (\ref{eq:inner3}) has also been discussed
extensively from other perspectives \cite{bran2,bran}.  
In these works, the inner product is defined
on $[0, \infty)$ rather than a finite interval, with the consequent
divergence (e.g., for the inner product between two QNMs each growing exponentially
at infinity) handled either (a) by a regulating factor $\exp(-\epsilon x^2)$,
$\epsilon \rightarrow 0^+$, (b) analytic continuation in the wavenumber $k$,
or (c) complex rotation in the coordinate $x$.  Each of these procedures has
its limitations; in contrast, (\ref{eq:inner3}) makes no
reference to the outside or bath, and is computationally convenient and 
manifestly finite.

Under this bilinear map, $H$ is symmetric:
$(\Psi , H \Phi) = (\Phi, H \Psi)$,
which follows very simply from $DH = H^{\dagger}D$.
This key property is analogous to the
hermiticity of $H$ for conservative systems.  It is nontrivial,
in that surface terms that arise in the integration by parts are
exactly compensated by the surface terms in (\ref{eq:inner3}).
This symmetry property leads, in the usual
way, to the orthogonality of non-degenerate eigenfunctions.

The completeness relation (\ref{eq:comp1})
is a dyadic equation.  Its
$(1,2)$ and $(1,1)$ components lead to the sum rules \cite{fn4}

\begin{eqnarray}
\sum_n \frac{f_n(x)f_n(y)}{2\omega _n}&=&0
\nonumber \\
\sum_n \frac{1}{2}f_n(x)f_n(y)\rho(x)&=&i\delta (x-y)
\label{eq:comp4}
\end{eqnarray}

\noindent
for $x, y \in I$, which have been derived and discussed extensively \cite{twocomp}.

The completeness and orthogonality relationships establish
the QNMs as a BB, and moreover allow
the time evolution to be solved as
$|\Phi (x,t)\rangle = \sum_n a_n e^{-i\omega _nt} \, |F_n\rangle$,
where
$a_n = \langle G_n|\Phi (x,0)\rangle / (2\omega _n)$.
This is a {\em discrete}\/ and {\em exact}\/ representation of the dynamics,
even though $I$ is open to an infinite universe
with a continuum of states.  Completeness is not
proved in most other applications of NHHs
to physical systems. 

\noindent {\bf Perturbation theory}

These notions allow much of the standard 
formalism in
quantum mechanics to be carried over.  As one example consider
time-independent perturbation theory. Let $\rho _0(x)^{-1}$ be changed to $%
\rho (x)^{-1}=\rho _0(x)^{-1}\left[ 1+\mu V(x)\right] $, where $|\mu| \ll 1$ 
$V(x)$ has support in $I$. 
Then the perturbation to the eigenvalues and eigenfunctions
can be written in the standard Rayleigh-Schr\"{o}dinger form,
in terms of a {\em discrete}\/ series \cite{twocomp}.
These formulas, though superficially identical with textbook
formulas for conservative systems, are
nontrivial in two ways.  First, the perturbative
formulas apply to {\em complex}\/ eigenvalues.
Second, the use of resonances
implies that there is no ``background", and expressing the corrections 
in terms of discrete modes also means that the small
parameter of expansion is $\mu / |\Delta \omega| \sim \mu a / \pi $,
which would not have been apparent in terms
of the states of the continuum. 

The derivation of these results simply follows
the conservative case
(everywhere replacing inner products by the bilinear
map $(\Psi, \Phi)$), and need not be repeated.

\noindent {\bf Discussion}

We have established an exact correspondence
between phenomenological NHHs and waves in a class of open systems.
This relationship provides a well-founded realization of NHHs.
Because we start with a hamiltonian system and remove
the bath degrees of freedom without approximations,
these open systems can be second-quantized \cite{sq}.
In other words, one can discuss photons in open cavities using BBs,
which makes this class of examples unique and interesting.
The relationship also places these open systems
into a well-known and convenient framework.  Thus, the linear space structure,
orthogonality and completeness can all be derived naturally,
by transcribing usual derivations for conservative systems and everywhere
replacing the inner product by $(\Psi, \Phi)$.  

The formalism discussed here also applies to the Klein-Gordon equation
with a potential $V(x)$ \cite{kg}, which applies, among other things, to
linearized gravitational waves propagating away from a black hole.
The first-order perturbation result for the QNM frequencies has been used
to understand the shifts in the gravitational wave frequencies when a black
hole is surrounded by an accretion shell \cite{dirt}.

The wave equation discussed here may be regarded as 
a physical realization
of BBs for open systems.  Many
other inequivalent realizations arise when
one considers outgoing waves in a spherically
symmetric 3-d system; each angular momentum $l$ leads
to realizations in which the surface terms
in the inner product involves $l$ radial derivatives
\cite{out}.

However, the entire formalism refers to systems
described by second-order differential equations,
so that two sets of initial data, namely $\phi$ and
${\hat \phi}$, are required, and the outgoing
condition is expressed as a constraint between them.
The formalism does {\em not} apply in its entirety
to systems described by first-order differential equations, 
e.g., $\alpha$-decays described by the Schr\"{o}dinger equation
with Gamow boundary condition.
In any event, the Schr\"{o}dinger equation formally
gives unbounded signal speeds and does not possess
outgoing and incoming sectors related by time reversal;
thus the concept of outgoing waves is actually quite different.
Nevertheless, if one is interested only in frequency domain
problems, e.g., eigenvalue problems and time-independent
perturbation theory, then the formalism
survives even in this case. This is most easily appreciated
by starting with the Klein-Gordon equation and simply
relabelling $\omega^2 \mapsto \omega$.

Using $(\Psi,\Phi)$ rather than the equivalent form
$\langle D \Psi | \Phi \rangle$ allows all reference to $D$ to be avoided.
However, $(\Psi,\Phi)$ is a bilinear map (rather than being
linear in the ket and conjugate linear in the bra).  This property is quite general, since
$D$ is conjugate linear.  But in most applications of the inner product
(e.g., for projections), it does not matter whether the map is linear
or conjugate linear in the bra; this is why results from conservative
systems can be carried over.  The only property that is lost is the positivity
of $(\Phi,\Phi)$, which is unsurprising
for a dissipative system.  Thus it is useful to think of the states of quantum dissipative
systems as vectors in a linear space $W$ endowed with such a bilinear map,
which is the generalization of the notion of a Hilbert space.
Time-evolution is then generated by an operator $H$ which is symmetric.

The open systems
described here are genuinely dissipative, with
$\mbox{Im } \omega_n < 0$.  This contrasts with some
models with NHHs which are nevertheless conservative \cite{sun,wu}.
For infinite-dimensional NHH models, completeness of the BB
is usually {\em assumed}, but difficult to prove.  Through these
wave systems, we have provided explicit examples where completeness
can be proved (if the discontinuity and ``no tail" conditions are met),
as well as examples where the basis is not complete (if these conditions
are not met).  These should also be useful in 
furthering understanding of NHH 
models.

We thank C. K. Au, E. S. C. Ching, S. Y. Liu 
and A. Maassen van den Brink for discussions.
This work is supported in part by the Hong Kong Research Grants Council
(Grant no. 452/95P). The work of WMS is also supported by the US NFS (Grant
no. PHY 96-00507), and by the Institute of Mathematical Sciences of The
Chinese University of Hong Kong.  The work of CPS at The Chinese University
of Hong Kong is also supported by a C. N. Yang Fellowship.


\end{multicols} 
 

\begin{references}

\bibitem{diss}
P. Ullersma, Physica {\bf 32}, 27 (1966);
R. P. Feynman and F. L. Vernon, Ann. Phys. {\bf 24}, 118 (1963);
P. S. Riseborough, P. Hanggi and U. Weiss, Phys. Rev. A. {\bf 31}, 471 (1985);
A. O. Calderira and A. J. Leggett, Ann. Phys. (N.Y.) {\bf 149}, 374 (1983);
L. H. Yu and C. P. Sun, Phys. Rev. {\bf 49}, 592 (1994).

\bibitem{wong}
J. Wong, J. Math. Phys. {\bf 8}, 2039 (1967).

\bibitem{lath}
L. Lathouwers, in {\em Quantum Science Methods and Structures},
J.-L. Calais {\em et al.} Eds. (Plenum Press, N. Y., 1976). 

\bibitem{rott}
E. Persson, T. Gorin and I. Rotter,
Phys. Rev. E {\bf 54}, 3339 (1996).

\bibitem{faisal}
F. H. M. Faisal and J. Moloney, J. Phys. B. {\bf 14}, 3603 (1981);
H. Baker, Phys. Rev. A {\bf 30}, 773 (1984).

\bibitem{dattoli}
G. Dattoli, A. Torre and R. Mignani, J. Phys. A {\bf 23}, 5795 (1990).

\bibitem{sun93}
C. P. Sun, Physica Scripta {\bf 48}, 393 (1993).

\bibitem{others}
V. F. Weisskopf and E. P. Wigner, Z. Phys. {\bf 63}, 54 (1930);
{\bf 65}, 18 (1930);
H. Feshbach, Ann. Phys. {\bf 5}, 357 (1958); {\bf 19}, 287 (1962);
L. Fonda and R. G. Newton, Ann. Phys. {\bf 10}, 490 (1960);
Y. Aharonov, {\it et al.},
Phys. Rev. Lett. {\bf 77}, 983 (1996).

\bibitem{fn2}
The latter condition is appropriate for waves
emitted at some finite time in the past, and differs
from usual discussions of resonances.

\bibitem{string}
H. Dekker, Phys. Lett. A {\bf 104}, 72 (1984);
Phys. Lett. A {\bf 105}, 395 (1984);
Phys. Lett. A {\bf 105}, 401, (1984);
Phys. Rev. {\bf 31}, 1067 (1985);
H. M. Lai, P. T. Leung and K. Young, Phys. Lett. A {\bf 119}, 337 (1987).

\bibitem{lang}
R. Lang, M. O. Scully and W. E. Lamb, Phys. Rev. A {\bf 7}, 1788 (1973).

\bibitem{price}
R. H. Price, and V. Husain, Phys. Rev. Lett. {\bf 68}, 1973 (1992).

\bibitem{kg}
E. S. C. Ching, {\it et al.}, Phys. Rev. Lett. {\bf 74}, 2414 (1995);
Phys. Rev. Lett. {\bf 74}, 4588 (1995).

\bibitem{chand}
S. Chandrasekhar, {\em The Mathematical Theory of
Black Holes} (Univ. of Chicago Press, 1991).

\bibitem{abram}
Abramovici, A. A., {\em et al.}, Science {\bf 256}, 325 (1992).

\bibitem{lly1}
P. T. Leung, {\it et al.}, Phys. Rev. A {\bf 49}, 3057 (1994);
Phys. Rev. A {\bf 49}, 3068 (1994);
Phys. Rev. A {\bf 49}, 3982 (1994).

\bibitem{fn1}
These conditions allow the freedom $D \mapsto cD$ for any complex number $c$.
We fix the normalization by
$D^{\dagger}D=1$, and leave the phase convention to be specified later.

\bibitem{fn5}
The denominators in  (\ref{eq:comp1}) and
(\ref{eq:hrep}) may vanish.
This can be handled as the limit of nearby QNM
frequencies coalescing, and is seen to be exceptional
on account of level repulsion (in the real parts of the
freqeuncies).

\bibitem{chu}
See, e.g., C. Chu, {\em An Introduction to Wavelets}
(Academic Press, 1992).

\bibitem{stanton}
J. F. Stanton and R. J. Bartlett, J. Che. Phys. {\bf 98}, 7029 (1993).

\bibitem{morse}
See, e.g., P. M. Morse and H. Feshbach,
{\it Methods of Theoretical Physics} (McGraw-Hill, 1953).

\bibitem{cheng}
Y.-J. Cheng, C. G. Fanning and A. RE. Siegman,
Phys. Rev. Lett. {\bf 77}, 627 (1996).

\bibitem{mis}
T. Sh. Misirpashaev, P. W. Brouwer and C. W. J. Beenakker,
Preprint chao-dyn/9706026 (1997).

\bibitem{fox}
J. L. Remo, Applied Optics {\bf 28}, 6561 (1995).

\bibitem{sun}
C. P. Sun, Chin. Phys. Lett. {\bf 6}, 481 (1989).

\bibitem{wu}
Z. Y. Wu, Phys. Rev. A {\bf 40}, 6852 (1989).

\bibitem{twocomp}
P. T. Leung, S. S. Tong and K. Young, J. Phys. A {\bf 30}, 2139 (1997);
J. Phys. A {\bf 30}, 2153 (1997). 

\bibitem{bj}
See, e.g., J. D. Bjorken and S. D. Drell, 
{\it Relativistic Quantum Mechanics} (McGraw-Hill, New York, 1964).

\bibitem{zel}
Ya. B. Zeldovich, Zh. Eksp. Teor. Fiz. {\bf 39}, 776 (1960)
[Sov. Phys. - JETP {\bf 12}, 542 (1961)].

\bibitem{per1}
H. M. Lai, {\it et al.}, Phys. Rev. A {\bf 41},
5187 (1990);
Phys. Rev. A {\bf 41}, 5199 (1990).

\bibitem{bran2}
N. Elander and E. Br\"{a}das, p. 541 in
{\em Lecture Notes in Physics Vol. 325}, 
N. Elander and E. Br\"{a}das, Eds. 
(Springer-Verlag, N. Y. 1989). 

\bibitem{bran}
E. Br\"{a}ndas, p. 149 in
{\em Dynamics During Spectroscopic Transitions},
E. Lippert and J. Macomber Eds. (Springer-Verlag, N. Y. 1995).

\bibitem{dirt}
P. T. Leung, {\it et al.}, Phys. Rev. Lett., {\bf 78}, 2894 (1997).

\bibitem{fn4}
The analogous sum rule arising from the $(2,1)$ component 
is not convergent.

\bibitem{sq}
A. Maassen van den Brink, {\it et al.},
(in preparation).

\bibitem{out}
C. Y. Chong {\em et al.} (in preparation).

\end{references}
\end{document}